\documentclass[pdflatex,sn-mathphys-num]{sn-jnl}

\usepackage{graphicx}%
\usepackage{multirow}%
\usepackage{amsmath,amssymb,amsfonts}%
\usepackage{amsthm}%
\usepackage{mathrsfs}%
\usepackage[title]{appendix}%
\usepackage{xcolor}%
\usepackage{textcomp}%
\usepackage{manyfoot}%
\usepackage{booktabs}%
\usepackage{algorithm}%
\usepackage{algorithmicx}%
\usepackage{algpseudocode}%
\usepackage{listings}%

\usepackage{mathtools} 

\usepackage{todonotes}

\usepackage{tipa}

\usepackage{lineno}
\usepackage{etoolbox}

\AtBeginEnvironment{equation}{\linenomath}
\AtEndEnvironment{equation}{\endlinenomath}

\AtBeginEnvironment{align}{\linenomath}
\AtEndEnvironment{align}{\endlinenomath}

\theoremstyle{thmstyleone}%
\theoremstyle{thmstyletwo}%

\theoremstyle{thmstylethree}%

\begin{document}

\title[Software Entropy: A Statistical Mechanics Framework for Software Testing]{Software Entropy}
\subtitle{A Statistical Mechanics Framework for Software Testing}


\author*[1,2]{\fnm{Jerónimo} \sur{Fotinós}}\email{jerofotinos@gmail.com}

\author[3,4]{\fnm{Juan B.} \sur{Cabral}}\email{jbcabral@unc.edu.ar}

\affil*[1]{\orgdiv{Condense Matter Group, Department of Physics}, \orgname{Faculty of Mathematics, Astronomy, Physics, and Computation (FaMAF), National University of Córdoba (UNC)}, \orgaddress{\street{Medina Allende s/n}, \city{Córdoba}, \postcode{5000}, \state{Córdoba}, \country{Argentina}}}

\affil[2]{\orgdiv{Enrique Gaviola Physics Institute (IFEG)}, \orgname{National Scientific and Technical Research Council (CONICET)}, \orgaddress{\street{Medina Allende s/n}, \city{Córdoba}, \postcode{5000}, \state{Córdoba}, \country{Argentina}}}

\affil[3]{\orgdiv{Grupo de Innovación y Desarrollo Tecnológico}, \orgname{Comisión Nacional de Actividades Espaciales (GVT-CONAE)}, \orgaddress{\street{Ruta Provincial C45, km 8}, \city{Falda de Cañete}, \postcode{5187}, \state{Córdoba}, \country{Argentina}}}

\affil[4]{\orgname{Consejo Nacional de Investigaciones Científicas y Técnicas (CONICET)}, \orgaddress{\street{Godoy Cruz 2290}, \city{Ciudad Autónoma de Buenos Aires}, \postcode{C1425}, \country{Argentina}}}


\abstract{The notion of \emph{software entropy} is often invoked to describe the tendency of software systems to become increasingly disordered as they evolve, yet existing approaches to quantify it are largely heuristic. In this work we introduce a formal definition of software entropy grounded in statistical mechanics, interpreting test suites as executable specifications, that is, as macroscopic constraints on the space of possible program implementations. Within this framework, mutation analysis provides a practical approximation of the locally accessible microstate space, allowing entropy-related quantities to be estimated empirically. We propose metrics that quantify how test suites restrict program space, including an information-weighted measure of the distribution of constraint power across tests. Applying these ideas to a real-world project, we show how test suites reduce software entropy and how information weights reveal structural differences in the contribution of individual tests that traditional metrics such as code coverage fail to capture.}

\keywords{software entropy, testing, statistical mechanics, software quality}



\maketitle

\section{Introduction}
\label{section:intro}

\begin{quote}
All repairs tend to destroy the structure, to increase the entropy and disorder of the system. [\ldots] Systems program building is an entropy-decreasing process, hence inherently metastable. Program maintenance is an entropy-increasing process, and even its most skillful execution only delays the subsidence of the system into unfixable obsolescence.
\end{quote}

This observation, articulated by Fred Brooks in \textit{The Mythical Man-Month}~\cite{brooks1974mythical}, draws on earlier work by Lehman and Belady~\cite{belady1971programming}, who studied the evolution of large operating systems and arrived at these conclusions from what Brooks himself describes as ``a statistical mechanical model''. The analogy with thermodynamics is explicit: software systems, like physical systems, tend toward disorder. Yet, despite this early recognition of the connection between software evolution and statistical mechanics, the analogy has remained largely metaphorical.

This intuition of ``disorder'' resonates strongly with software development practitioners \citep{Hanssen2010SoftwareEntropy}, who routinely observe that software systems tend to become progressively harder to maintain, more error-prone, and increasingly resistant to modification as they grow and evolve. Yet, despite the ubiquity of this intuition and its apparent explanatory power, it has not been formalized in a manner that is both rigorous and compatible with statistical physics. To date, notions of entropy in the software lifecycle have largely remained at the level of metaphor, drawing loose analogies with the entropy defined in statistical mechanics rather than deriving from it in a principled way. Nevertheless, several lines of work have emerged over time that attempt to operationalize related ideas.


A first category applies Shannon's information entropy to artifacts related to software, but not to the software itself. \citet{Hassan2009} introduced the concept of \textit{change entropy}, which measures the dispersion of code changes across the files of a project, showing that more dispersed changes are better predictors of defects than traditional metrics. This approach was subsequently extended by \citet{Canfora2014}, who investigated how refactoring activities tend to reduce entropy while the involvement of multiple developers increases it, and by \citet{Kaur2017}, who incorporated entropy-based metrics into machine learning models for defect prediction. Along a different line, \citet{Niepostyn2023} applied entropy-like measures to the frequency of identifier occurrences in UML diagrams in order to assess design consistency. Other works have employed information entropy in adjacent domains: \citet{Wojnowicz2016} computed byte-level entropy of binary files to detect malware, while \citet{Fan2022} used similar techniques to identify distributed denial-of-service attacks in network traffic. While these approaches constitute valuable contributions, they all rely on Shannon entropy as a descriptive statistic, without establishing a true correspondence with a statistical ensemble that would endow entropy with a physical interpretation—the source of its deeper explanatory power.


A second category comprises empirical, ad-hoc definitions that attempt to quantify software disorder directly \citep{Trienekens2009}. These proposals, however, typically rest on arbitrary criteria lacking rigorous theoretical justification, combining heterogeneous metrics through functional forms that are not derived from any underlying fundamental principle.


Strikingly absent from both categories is any rigorous connection to statistical mechanics—the theoretical framework in which entropy was originally defined as a measure of the number of accessible microstates compatible with given macroscopic constraints. Boltzmann's formula $S = k \log W$, a foundational equation of statistical physics, has never been systematically applied to the space of possible programs. Moreover, no existing work establishes a formal link between software entropy and software testing, despite the fact that testing constitutes the primary mechanism by which program behavior is specified and verified. This omission is particularly surprising given that test suites define, in a very concrete sense, the observable properties of a software system—analogously to how temperature and pressure define the macrostate of a gas.


While more rigorous methods for specifying software do exist, testing has consistently remained the preferred approach in real-world projects, serving not only as a specification mechanism but also as a safeguard against defects \citep{brooks1987no}. Software testing spans a wide spectrum, ranging from unit tests, which validate individual components in isolation, through integration tests, which assess how different parts of a system interact, to test-driven development (TDD) \citep{beck2003test}, which fully embraces the idea of specifying software behavior through tests written prior to the implementation that is meant to satisfy them. However, there are several other distinctions unrelated to the previous classifications. For instance, one may distinguish between: example-based tests, which verify that a known input produces the expected output, and property-based tests \citep{claessen2000quickcheck}, which validate that a given property holds across many randomly generated inputs. There are also quantitative metrics such as code coverage \citep{pathy2015review}, which measures the fraction of the source code exercised by the test suite.
Beyond that,  mutation testing \citep{demillo2006hints} offers a systematic way of introducing perturbations into the source code to assess whether the existing tests are capable of detecting the injected faults, thereby evaluating the robustness of the testing process itself.
It is also worth keeping in mind some other testing techniques, such as: genetic testing~\citep{mcminn2004search}, which explores the space of inputs to automatically unveil patterns or properties unspecified but satisfied by the software; and fuzzy testing~\citep{miller1990empirical}, which subjects programs to random or problematic inputs in order to detect new or unexpected errors.


In this work, we introduce a theoretically grounded definition of software entropy based on the formalism of statistical mechanics. We propose that microstates correspond to concrete implementations of source code, while macrostates are defined by the set of tests that a program satisfies. Within this framework, software entropy quantifies the uncertainty about an implementation given that it satisfies a specification, operationalized here through a test suite. This formulation not only provides a solid theoretical foundation for long-standing developer intuitions, but also establishes a direct connection between the quality of a test suite and entropy reduction, thereby opening new perspectives for the evaluation and improvement of testing practices.

In addition, we present an empirical analysis of real-world software systems, together with a tool that allows to quantify an upper bound on the entropy of a given implementation, based on its test suite and the technique of mutation testing.

\section{Theoretical Formulation} \label{section: theoretical_formulation}
\subsection{Program Space: micro- and macrostates}
We begin our formulation by considering the program in question as a sequence of $L_{code} \in \mathbb{N}$ characters, each of which has $N_{char} \in \mathbb{N}$ possible values. Note that such a sequence actually includes all programs of \textit{up to} $L_{code}$ characters, for the last ones can be, for instance, blank spaces. For the sake of simplicity, we can consider that our entire code base is contained in a single file, without loss of generality. Then, our program is defined by a string of length $L_{code}$. This string might be thought of as a point in the lattice $\mathbb{Z}^{L_{code}}$ by just assigning a coordinate to each character. In fact, all possible programs lie in a subset of this lattice, given by allowing each of the $L_{code}$ coordinates to take values only in $[1, N_{char}]$. We may call such a subset $\mathbb{Z}_{[1, N_{char}]}^{L_{code}}$.

The first thing to consider is that most of the points within this subset will not comply with the programming language syntax. Let us call $\mathbb{V} \subset \mathbb{Z}_{[1, N_{char}]}^{L_{code}}$ the ``syntactic'' subset, i.e., the set of programs that do comply with syntax. Each point in $\mathbb{V}$, that specifies the source code of a program, corresponds to a \textit{microstate} from a statistical mechanical perspective. It would be analogous to a simultaneous specification of the positions and velocities for all particles of an ideal (classical) gas. To continue with this analogy, it would be convenient to ask ourselves what would be the \textit{macrostate} in this context. That is, we want to characterize a general ``macroscopic'' property of our program, without concerning ourselves with the details of the code. Since this coarse-graining is to be made in a way that describes relevant behavior, the natural type of restriction in this context is a semantic one. This suggests the use of the properties enforced by tests as the macroscopic restrictions. These would be the quantities analogous to temperature or pressure in a gas.

In order to formalize this notion, let $\{t_1, \dots, t_m\}$ be the set of statements (booleans) enforced by the tests, i.e., the propositions about our program that are true if our tests pass. These semantic restrictions are imposed properties of the system as a whole, and we may thus refer to them as the observable properties defining the macrostate. We can then define the ``semantic'' set $\mathbb{P} \subset \mathbb{V}$ of all microstates compatible with the observed macrostate as $$ \mathbb{P} = \{ x \in \mathbb{V} : (\land_{i=1}^m t_i)x = 1 \} $$ where $t_i$ is an operator that maps program strings into booleans that state whether the program fulfills the property enforced by the test. Thus, a concrete version of the program that passes the tests is a microstate $p \in \mathbb{P}$.

\subsubsection{Defining Software Entropy}
Now that we have our program space (or configuration space, in the language of mechanics), composed of microstates $p \in \mathbb{P}$ with macrostate $(t_1, \dots, t_m)$, we proceed to define the software entropy. We will initially do this from the perspective of information theory.

We could think of programmers as data emitters that produce $p \in \mathbb{P}$. More concretely, let $P$ be a discrete random variable that may have any outcome $p$ in the sample space $\mathbb{P}$, with probability distribution $\mu:\mathbb{P}\to[0, 1]$. Then, the information entropy $H$ would be
\begin{align}
    H(P) = -\sum_{p \in \mathbb{P}} \mu(p) \log \mu(p), \label{eq: information entropy of software}
\end{align}
where the sum is over all $p \in \mathbb{P}$, and the base of the logarithm determines the units of information.

This quantity is interpreted as the information we would gain if told the microstate while knowing the macrostate. If our tests are permissive and allow for many microstates to belong to $\mathbb{P}$, it would be hard to guess the outcome of a realization of $P$. Thus, we would actually gain information if told the precise microstate (the program), despite knowing the tests that the program complies with. On the other hand, consider the limit case of such restrictive tests that there is only a single possible program $p^{*}$ in $\mathbb{P}$. In that case, $H(P) = - \mu(p^{*}) \log \mu(p^{*}) = - \log(1) = 0$, reflecting that we would not get any additional information if told the microstate, because there is already a single program compatible with the macrostate $(t_1, \dots, t_m)$. This is similar to what happens in the case of a Bose–Einstein condensate. The macrostate in that case would be given by $T \to 0$, and no additional information would be gained by being told the microstate, since we know that the only possible microstate at that temperature is the fundamental state $| n_0 \rangle$ (the state of minimum energy). Of course, this analogy only intends to clarify the interpretation of the proposed definition of entropy by comparing it to a well-known use case. For the sake of simplicity, we have not mentioned the possibility of degeneracies in the fundamental state $| n_0 \rangle$, since conceptually the point still holds. Indeed, such degeneracies not only happen in physical systems, but can also occur in our program space: there may be two concrete program implementations, within the established program size, that satisfy a complete, rigorous specification (we only test behavior). In such a case, the entropy minimum will be greater than zero, but nothing else will have changed. Moreover, this modification would be inconsequential, as state functions are usually defined up to an additive constant. This is perfectly compatible with the fact that for both physical systems and our program behavior case, we are usually interested in differences in these quantities between states. For physical systems, we usually aim at calculating entropy or internal energy differences. For programs, we may be interested in the decrease in entropy after a sprint in which the testing team added new tests for the program to pass.

\subsection{An Upper-Bound on the Entropy} \label{section: Entropy Upper-Bound}
A complication that arises in this formulation is that the probability distribution $\mu$ will generally not be accessible to us. This is so, since it reflects the likeliness of programmers coding $p$ under the restriction of passing tests $(t_1, \dots, t_m)$. However, we do have access to an upper bound of $H(P)$ from which we can extract some insights. This upper bound, which we will call $S$, is given by the case in which all outcomes are equiprobable. In that case, eq. \eqref{eq: information entropy of software} reduces to
\begin{align}
    S = \log W
    \label{eq: entropy upper-bound}
\end{align}
where $W$ is the cardinality of $\mathbb{P}$. Note that eq. \eqref{eq: entropy upper-bound} is the very well-known Boltzmann formula for the statistical mechanical entropy of a thermodynamical system, with the exception perhaps of a multiplicative constant. In statistical thermodynamics, this formula is derived from the Gibbs entropy formula, which has the same form as \eqref{eq: information entropy of software}, in a similar manner. For physical systems in equilibrium, the assumption of equiprobability usually holds since for constant total energy, there's no reason to privilege one state over another. This is known as the \textit{equal a priori probability postulate} \cite{Tolman1938}: phase points for a given system are equally likely when they correspond equally well with the knowledge we have of the system. This means that the system can be found with equal probability in any microstate consistent with that knowledge.

There are many arguments in favor of assuming equiprobability. In physical systems, the main argument is that the ergodic hypothesis usually holds for systems like the ideal gas. A system is said to be ergodic when, during its evolution, it is able to get to every state allowed by external restrictions. This is why in statistical mechanics we are able to replace temporal averages over a realization or trajectory of the actual system, by ensemble averages over all accessible configurations. Although we could conceive programmers modifying the code under the same tests, describing a potentially ergodic trajectory, this is not an appropriate assumption in our case.

Another argument consists of invoking the principle of maximum entropy, which states that the best probability distribution given prior data is the one that maximizes the entropy. In our case, this could be interpreted as taking the simplest non-informative prior and then getting the least biased probability distribution that is compatible with the data, under a Bayesian probability approach to inference.

\subsection{Entropy and the Life of a Program}
During the life of our program, we expect its size to increase, i.e., we would have $\Delta L_{code} \geq 0$. If no additional tests are imposed, this would mean that $\Delta W \geq 0$. Whereas it's reasonable to think that this would imply that the entropy would grow, i.e., $\Delta H(P) \geq 0$, we have no way of proving this for the general case. Nonetheless, we can point out that the upper-bound $S$ do in fact grow, thus giving $ \Delta S \geq 0$. This means that, while the strict proof would depend on the particular form of the distribution $\mu$, which remains inaccessible, entropy growth is to be expected when the code base increases in size.

Moreover, since a greater number of programs passing our tests increases the probability of unexpected behavior, higher entropy is linked with a greater probability of such behavior. This is not surprising if we think of it from the perspective of information theory: there is now more information to be gained from looking at the program while knowing that it passes the tests.

Since our code is not a closed system, entropy growth with time is not a strict law as the one that gives the arrow of time for isolated physical systems. Under this definition, software entropy characterizes the expected surprise under the restrictions of the tests. This can be reduced in many ways, namely, by incorporating new tests. As mentioned before, the limit case is given by, for instance, an extreme form of Test-Driven Development (TDD) under which there's a single, well-defined program that passes the tests. That is, $W=1 \implies S=0$ and since $0 \leq H(P) \leq S = 0 \implies H(P) = 0$.

\subsection{A First Step Towards Modeling Unexpected Behavior}
In principle, not all unexpected behavior would constitute a bug. We could have innocuous differences in programs that do not really pose a problem, constituting what we will call ``equivalent mutants'' from now on. In order to provide a rationale regarding how entropy and damaging behavior might be linked, let us consider the following. Assume that there exists a partition of $\mathbb{P}$ given by $\{\mathbb{P}_{+}, \mathbb{P}_{-}\}$, of respective cardinalities $W_+$ and $W_-$, the first one containing innocuous variations of the code, and the second, malicious ones. For instance, elements on $\mathbb{P}_{+}$ could differ in performance, do extra assertions, differ in dependencies, handle contexts in different ways, etc. On the other hand, elements of $\mathbb{P}_{-}$ would contain variations that result in undesired behavior.

In principle, what we would want is for our program to belong to $\mathbb{P}_{+}$. Of course, we cannot compute the precise probability of that being the case, but under the assumption of maximum entropy, we could say that the probability of the software being defective is proportional to the fraction of microstates that belong to $\mathbb{P}_{+}$, which is $f=W_{+}/W$. Under this framework, we conjecture that as the code base grows, this fraction would become smaller. This would be in accordance with the intuition of some developers that software systems become more buggy, disordered, and hard to modify as they grow.

\medskip

\noindent\textbf{Undecidability of Membership in $\mathbb{P}_{+}$ and $\mathbb{P}_{-}$.}
It is important to highlight that, unlike the test-based macrostate $\mathbb{P}$, the
partition $\{\mathbb{P}_{+},\mathbb{P}_{-}\}$ appeals to semantic criteria describing whether
a program exhibits benign or harmful behavior on \emph{all} possible inputs. Such
properties depend on the extensional behavior of the program and therefore fall within
the scope of Rice's theorem, which states that every non-trivial semantic property of
partially computable functions is undecidable in general. Consequently, membership in
$\mathbb{P}_{+}$ or $\mathbb{P}_{-}$ cannot be algorithmically determined for arbitrary
programs, even when the code length is fixed. This does not undermine the usefulness of
our framework: it simply clarifies why empirical probes like tests provide the only feasible means of approximating these sets.

\subsection{Testing and Software Entropy}
In all cases, putting in place more restrictive (non-trivial) tests reduces entropy. It decreases the likelihood of unintended behavior, as it augments our knowledge of the program, rendering the observable state more descriptive. To be more explicit, putting in place a new (non-redundant) test $t_{m+1}$, in addition to the previously existing restrictions $(t_1, \dots, t_m)$, makes the macrostate more specific, eliminating possible microstates of $\mathbb{P}$. Up to this point, we have considered this property $t_{m+1}$ as enforced by unitary or integration tests, without that distinction being important. Let us now consider other types of testing and their meaning under this formalism.

Beyond their operational role, testing techniques provide structural information about program space. Each class of tests contributes a different type of restriction, and therefore a different form of entropy reduction. In what follows, we examine mutation testing and how it probes the structure of $\mathbb{P}$. For a discussion on other types of testing, refer to Appendix \ref{appendix: types_of_testing}.

\subsubsection{Mutation Testing and Mutation Graph} \label{section: mutation_testing_and_graph}

Mutation testing produces syntactic variants $\tilde{p} \in \mathbb{V}$ of the implemented program $p_{\mathrm{impl}} \in \mathbb{P}$,
and then evaluates which of these ``mutants'' \textit{survive}, i.e., remain compatible with a given set of tests, which simply means checking if $\tilde{p}\in \mathbb{P}$.

It is important to note that the mutation operator induces a graph $G=(\mathbb{V},E_M)$ where edges connect programs related by a mutation step. The induced subgraph $G[\mathbb{P}]$ contains all syntactic variants that pass the tests. Basic graph-theoretic features---the number of connected
components, and the size of the largest component---provide a coarse picture of how
``fragile'' or ``robust'' membership to the current macrostate is: highly fragmented feasible regions
indicate that small syntactic perturbations tend to violate constraints. On the other hand, a robust basin indicates that tests are locally permissive, something that is generally undesirable. In this way, mutation testing allows us to identify how robust membership to $\mathbb{P}$ is in a neighborhood of $p_{impl}$.

\subsubsection{Evolving Test Suites and Entropy Reduction} \label{section: evolving_test_suites}

As stated in the previous section, mutation testing gives us a technique to probe for how restrictive a test suite is: the more fragile membership to $\mathbb{P}$ is, the better specified the program is. As it turns out, one may use mutation testing to assess the impact in software entropy of changes in the test suite across time.

Say that $p_{\mathrm{impl}}$ is the implemented program. Let $\mathcal{F}=\{T_j\}_{j=0,1,\dots}$ be an increasing (i.e. nested) family of test sets, with $T_0=\varnothing$ and $T_{j_1}\subset T_{j_2}$ for all $j_1<j_2$. In what follows, $\mathcal{F}$ should be interpreted as a set containing the evolution of a test suite (the test suite at time $j$ is $T_j$). Each $T_j$ induces a local feasible set
\[
M_j(p_{\mathrm{impl}}) := \{\,q\in M(p_{\mathrm{impl}}): t_k(q) = 1, \forall t_k \in T_j\,\},
\]
where $M(p_{\mathrm{impl}})$ is the collection of mutants produced by a chosen mutation operator that may introduce more than one change at a time. The local acceptance ratio of mutants allowed by $T_{j_1}$ when the macrostate is strengthened to $T_{j_2}$ is
\begin{align*}
    \rho_{j_1,j_2} \coloneqq \frac{|M_{j_2}(p_{\mathrm{impl}})|}{|M_{j_1}(p_{\mathrm{impl}})|},
\qquad j_2>j_1,
\end{align*}
and the corresponding reduction in the micro\-canonical entropy upper-bound is

\begin{align}
    \Delta S_{j_1,j_2}(p_{\mathrm{impl}})
= -\log \rho_{j_1,j_2}
= \log |M_{j_1}(p_{\mathrm{impl}})| - \log |M_{j_2}(p_{\mathrm{impl}})|.
\label{eq: entropy_loss}
\end{align}

Thus, performing mutation testing on a sequence of test suites allows us to quantify a \emph{local entropy loss} induced by the addition of new tests: more restrictive tests collapse a larger fraction of the mutation neighborhood and produce larger $\Delta S$. This type of analysis can also be performed on a test suite for which we do not know the history: By choosing the family $\mathcal{F}$ according to impact metrics (e.g., coverage impact, uniqueness, or redundancy), one obtains entropy-reduction curves $\Delta S_{0,j}$ that quantify how successive subsets of tests constrain program behavior. Although these curves probe only a neighborhood of $p_{\mathrm{impl}}$, they provide a practical proxy for the upper bound on how tests reduce local behavioral degrees of freedom.

\subsection{Local Test-Suite Metrics}
\label{section: local metrics}

While mutation score measures the overall fraction of mutants killed by a test suite, it does not reveal how the constraint power is distributed across individual tests. In particular, two test suites with identical mutation scores may differ substantially in how much each test contributes to restricting the space of admissible implementations.

To quantify this internal structure, we analyze the mutants uniquely eliminated by each test. For a test $t_i$, let $K_i$ denote the set of mutants killed only by $t_i$. The relative contribution of $t_i$ to the entropy reduction of the test suite can then be expressed through the \emph{information weight}

\begin{equation}
\alpha_i := \frac{|K_i|}{\sum_{j=1}^m |K_j|},
\label{eq: information_weight}
\end{equation}

which measures the fraction of uniquely eliminated mutants attributable to
that test. The weights $\{\alpha_i\}$ therefore quantify how the constraint
power of the test suite is distributed among its constituent tests.

A natural summary of this distribution is obtained by computing its Shannon
entropy. Normalizing by the maximum possible entropy $\log m$ yields the
\emph{Macrostate Tightness Index}

\begin{equation}
MTI_2 := \frac{-\sum_{i=1}^m \alpha_i \log \alpha_i}{\log m},
\label{eq: mti_2}
\end{equation}

which takes values in the interval $[0,1]$. Large values of $MTI_2$ indicate
that constraint power is evenly distributed across the test suite, while small
values reveal that only a few tests are responsible for most of the entropy
reduction.

Beyond providing the aggregate metric $MTI_2$, the weights $\alpha_i$ themselves
offer a finer-grained view of the role played by each test. In particular, they
identify tests that impose strong semantic constraints on the implementation
and those whose mutation-killing capacity is largely redundant with that of
other tests. This interpretation is explored empirically in
Section~\ref{sec: information_weights}. For an extended discussion on possible metrics, refer to Appendix \ref{appendix: local_metrics}.

\subsection{Time-Dependent Macrostates and Non-Equilibrium Analogy}

We believe a final remark on this ``static'' framework is pertinent. Software development is a dynamical process: both the implementation $p(t)$ and the set of
tests $T(t)$ evolve. Thus, the macrostate $\mathbb{P}(t)$ changes in time, and entropy may increase (code growth with no new constraints) or decrease (addition of tests or formal refinement). This resembles a driven, non-equilibrium system being steered through successive constraint surfaces, rather than an equilibrium ensemble with fixed admissible microstates. The local entropy-loss curve $\Delta S_{0,j}$ along a test-writing cycle is a concrete instance of this viewpoint. In this way, the possibility of such a generalization is reassuring, although it is beyond the scope of the present work and is unlikely to yield significant additional benefits.


\section{Empirical Analysis}
\label{section:empirical}

The direct computation of software entropy, for instance for the upper bound \eqref{eq: entropy upper-bound} defined in section \ref{section: Entropy Upper-Bound}, is computationally intractable.
In principle, a natural approximation would consist in performing a historical analysis of the repository using \textit{repository mining} techniques~\citep{hassan2008road}, reconstructing the evolution of both the test suites and the source code across successive commits.
However, this approach faces significant practical obstacles.

First, the computational cost of running mutation analysis across multiple historical versions of a project is prohibitive for repositories of realistic size~\citep{vercammen2018speeding}.
Second, and more fundamentally, common development practice—even within methodologies explicitly oriented toward testing such as TDD~\citep{beck2003test}—typically involves committing tests and functionality simultaneously.
That is, it is rare to find commits in a repository history that contain tests for code that has not yet been implemented; instead, each commit usually includes both the test and the implementation that satisfies it.
As a consequence, artificially reconstructing the sequence of macrostates $\{T_0, T_1, \ldots, T_m\}$ defined in Section~\ref{section: evolving_test_suites} would require manually disabling and enabling subsets of the code or the test suite for each analyzed version.
Such a procedure is not only tedious, but also introduces additional complications arising from inter-module coupling~\citep{offutt1993software}: disabling a test or a portion of the code can trigger cascading effects that are difficult to predict and control.

In light of these difficulties, we propose an alternative approach that allows us to approximate the entropy reduction induced by each individual test without reconstructing the repository history.
This approach is grounded in the mutation graph framework introduced in Section~\ref{section: mutation_testing_and_graph}. Rather than attempting to enumerate the entire (intractable) program space, we systematically explore the local neighborhood of the implemented program.

The procedure is as follows:
\begin{enumerate}
    \item A complete mutation analysis is executed on the project, yielding the full set of surviving mutants $M_{\text{total}}$ when all tests are enabled.
    \item For each test $t_i$ in the test suite:
    \begin{enumerate}
        \item Mutation analysis is executed with \textit{only} $t_i$ enabled, producing the set of surviving mutants $M_{\{t_i\}}$.
        \item Mutation analysis is executed with \textit{all tests except} $t_i$ enabled, producing the set of surviving mutants $M_{\overline{\{t_i\}}}$.
    \end{enumerate}
\end{enumerate}

This design makes it possible to quantify both the individual restrictive power of each test (by comparing $|M_{\{t_i\}}|$ with the total number of mutants) and its marginal contribution to the suite as a whole (by comparing $|M_{\overline{\{t_i\}}}|$ with $|M_{\text{total}}|$).
From these measurements, the metrics defined in Section~\ref{section: local metrics} can be computed, and entropy reduction curves can be constructed to characterize the quality of the test suite.

Clearly, carrying out this procedure systematically across multiple software packages using an ad-hoc script would be impractical.
Mutation analysis is computationally expensive, and the methodology described above requires executing it multiple times for each test in the suite.
Consequently, considerations such as the ability to resume interrupted runs, process parallelization, and efficient management of intermediate results become essential.
The natural solution to these requirements is the development of a dedicated tool capable of performing this type of analysis in a robust and reproducible manner.

\subsection{The Package: Yagua}
To address the requirements described above, we have developed \texttt{Yagua}\footnote{Yagua (or jagua; pronounced [\textipa{Za"\textlowering{G}wa}]) in Guaraní primarily means ``dog'', but in colloquial Argentine and Paraguayan usage it is often used as an exclamation expressing rejection or surprise in response to a false claim.}, a Python package that implements the mutation graph exploration methodology described in Section~\ref{section: mutation_testing_and_graph}. \texttt{Yagua} implements this methodology by systematically exploring the local neighborhood of a program in the mutation graph and managing the resulting test and mutation data for \texttt{pytest}-based projects.

The choice of \texttt{pytest}\footnote{\url{https://docs.pytest.org/}} as the testing framework is motivated by its widespread adoption within the Python community, while \texttt{Cosmic Ray}\footnote{\url{https://github.com/sixty-north/cosmic-ray}} was selected for the ease with which it integrates into automated workflows (the only comparable alternative in the ecosystem is \texttt{mutmut}\footnote{\url{https://github.com/boxed/mutmut}}).
In addition to mutation analysis, \texttt{Yagua} also collects code coverage metrics~\citep{pathy2015review} following the same selective test enabling and disabling scheme described in the previous section: for each test $t_i$, coverage is measured by executing only $t_i$, and complementarily by executing all tests except $t_i$.
This coverage information is particularly useful for estimating the impact of each test without running the full mutation analysis and, for example, for ordering test execution so as to obtain meaningful results more rapidly.
\texttt{Yagua} provides a command-line interface for executing these tasks, as well as interactive analysis capabilities over the collected results.
All collected data—tests, mutants, execution outcomes, coverage information, and derived metrics—are stored in a single database file, facilitating both subsequent querying and full reproducibility of the analysis.
A detailed description of the architecture, functionality, and use cases of \texttt{Yagua} will be presented in a future publication devoted exclusively to this tool.

\subsection{A Simple Case: Astrolign}

\texttt{Astroalign} is a Python module for astronomical image registration, that is, for aligning two images of the sky so that they match pixel by pixel.
Unlike approaches that rely on celestial coordinate metadata, \texttt{Astroalign} adopts a purely geometric strategy: it detects stars in both images, constructs triangles from the brightest detections, computes geometric invariants (such as ratios of side lengths that remain unchanged under rotation, translation, or scaling), and matches corresponding triangles across images to estimate the optimal affine transformation \citep{beroiz2020astroalign}.
At the time of writing, the package is widely used within the astronomical community, with approximately 5{,}575 monthly downloads according to PyPI Stats\footnote{\url{https://pypistats.org/packages/astroalign}}, and the associated publication has accrued 121 citations according to Google Scholar\footnote{\url{https://scholar.google.com/citations?view_op=view_citation&hl=es&user=0dAXzPYAAAAJ&citation_for_view=0dAXzPYAAAAJ:mB3voiENLucC}}.

The choice of \texttt{Astroalign} as a case study is deliberate.
One of the authors of the present work is also a coauthor of the \texttt{Astroalign} project, which allowed us to avoid the learning curve typically associated with understanding an unfamiliar codebase and, more importantly, to rely on expert knowledge regarding the intent behind each test and the software’s design decisions.
This level of insight is particularly valuable in the present context, where tests are interpreted as macroscopic constraints on the space of admissible implementations.

At the time of analysis, the project comprised 29 tests achieving a code coverage of 97.45\%.
For this study, 463 mutants were generated and analyzed.
All experiments were conducted using the \texttt{master} branch of the Git repository at commit
\href{https://github.com/quatrope/astroalign/commit/76077b2583601611f5f8f2d52be47ca481f77ab1}{\texttt{76077b2}}.

\subsection{Reducing Software Entropy} \label{sec: reducing_soft_entropy}

This section illustrates how the addition of tests reduces the software entropy introduced in Section~\ref{section: theoretical_formulation}. In the statistical-mechanical formulation presented in this work, the macrostate of a program is defined by the set of behavioral constraints imposed by the test suite, while microstates correspond to concrete implementations of the code that satisfy those constraints. Strengthening the macrostate by adding tests therefore reduces the number of admissible microstates and, consequently, the entropy.

To make this effect explicit, we simulate the progressive construction of a test suite. Let the complete suite be
\[
T=\{t_1,t_2,\dots,t_m\},
\]
and consider the nested sequence of partial suites
\[
T_i=\{t_1,t_2,\dots,t_i\}.
\]
Each $T_i$ defines a macrostate characterized by the set of implementations that satisfy the corresponding constraints. In practice, we approximate this set by exploring the local neighborhood of the implemented program on the mutation graph.

For each partial suite $T_i$, mutation testing is executed, and the number of surviving mutants $W_i$ is recorded. These surviving mutants represent program variants that remain compatible with the constraints imposed by $T_i$. Following Eq.~\eqref{eq: entropy upper-bound}, we estimate the corresponding entropy upper bound as
\[
S_i=\log W_i.
\]

The reduction in entropy associated with the addition of a new test can then be quantified by the entropy difference between successive macrostates,
\[
\Delta S_{i,i+1}=S_i-S_{i+1},
\]
which corresponds to the local entropy loss defined in
Eq.~\eqref{eq: entropy_loss}. In this sense, each new test introduces an additional semantic constraint that removes accessible microstates from the mutation neighborhood.

\begin{figure}
    \centering
    \includegraphics[width=\textwidth]{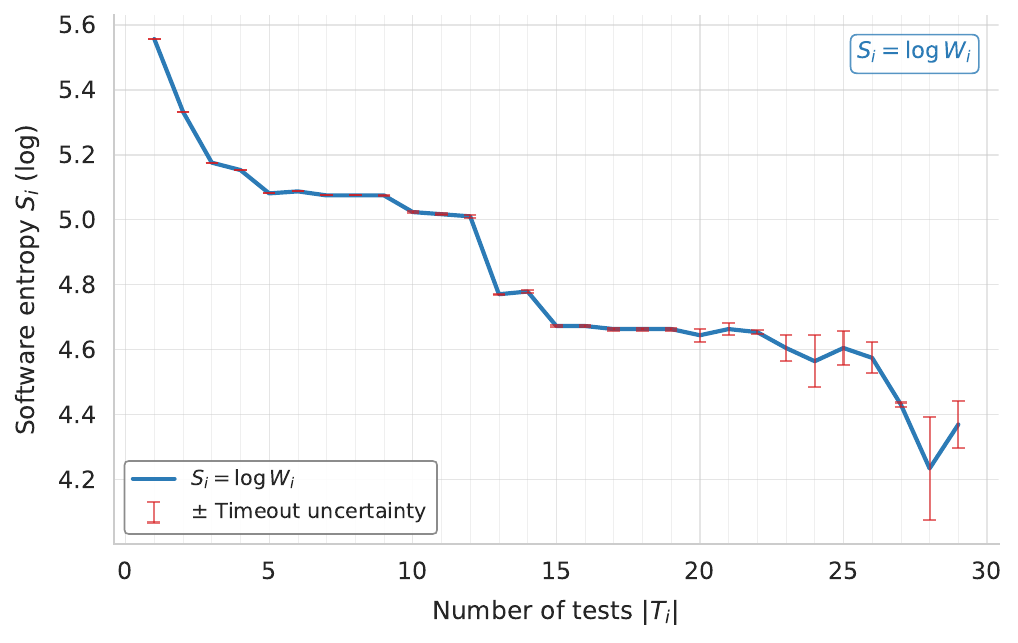}
    \caption{\emph{Software entropy $S_i$ as a function of the number of tests $|T_i|$.} The entropy upper bound is estimated on the mutation graph as $S_i=\log W_i$, where $W_i$ is the number of mutants that survive the partial test suite $T_i=\{t_j\}_{j=1}^i$. Error bars correspond to the number of mutation runs that resulted in timeouts, which are treated as uncertainties in the entropy estimate. The overall decreasing trend reflects the progressive reduction of accessible microstates as additional behavioral constraints are introduced by the test suite.}
    \label{fig: entropy_vs_tests}
\end{figure}

The resulting entropy trajectory is shown in Fig.~\ref{fig: entropy_vs_tests}, where the entropy upper bound $S_i$ is plotted
as a function of the test suite cardinality $|T_i|$. The horizontal axis may also be interpreted as a proxy for development time, for example, successive development iterations during which tests are added.

In an ideal deterministic setting, the number of surviving mutants could only decrease or remain constant as new tests are incorporated. Consequently, the entropy would be a monotonic non-increasing function of $|T_i|$. In practice, however, several factors introduce stochastic variability in the measurements.

First, either the code or the tests may contain components that rely on random number generators without fixed seeds. Such randomness may lead to occasional realizations in which a larger test suite appears to kill fewer mutants. Second, mutation testing may produce timeouts. By default, \texttt{CosmicRay} counts timeouts as killed mutants; from a theoretical perspective, however, this classification is inaccurate. For this reason we analyzed the mutation databases directly and counted timeouts separately from confirmed kills. Since a timeout may correspond either to a surviving or to a killed mutant, these cases are treated as uncertainties in the entropy estimate. 

Despite these sources of noise, the expected monotonic behavior would be recovered by averaging over an ensemble of identical experiments. While such an ensemble average would be conceptually straightforward from the perspective of statistical mechanics, it would be computationally prohibitive in practice. Nevertheless, even a single realization such as the one shown in
Fig.~\ref{fig: entropy_vs_tests} clearly demonstrates that strengthening the test suite reduces the number of admissible microstates and therefore lowers the
software entropy upper bound.

\begin{figure}
    \centering
    \includegraphics[width=\textwidth]{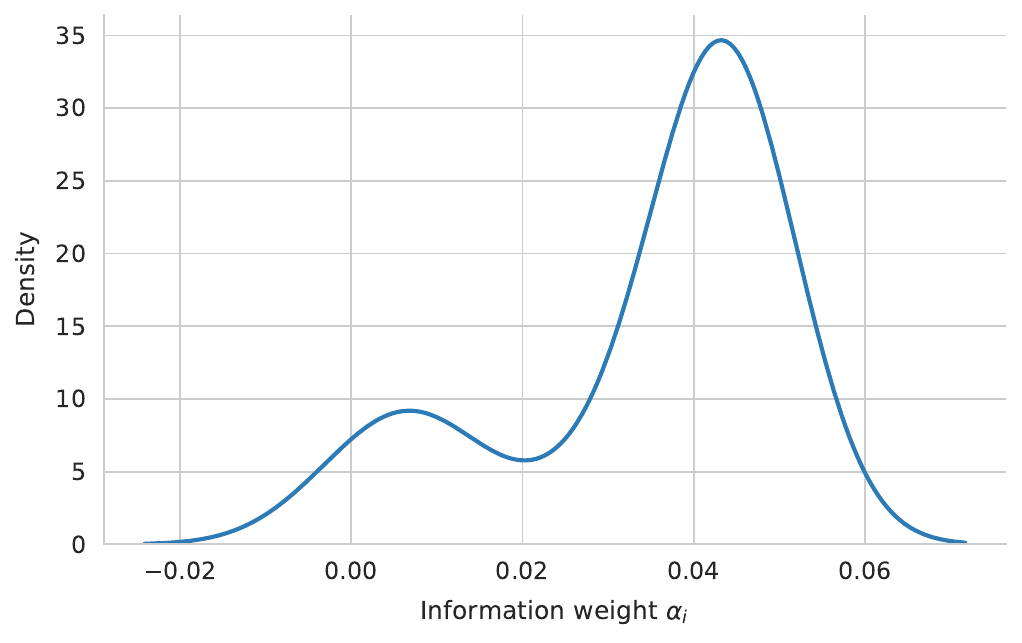}
    \caption{\emph{Distribution of information weights $\alpha_i$ across the test suite.}
Each weight is defined as $\alpha_i = |K_i| / \sum_j |K_j|$, where $K_i$ is the
set of mutants killed by test $t_i$ when run alone. The bimodal structure suggests
two distinct populations of tests: a minority with low individual restrictive power
and a dominant group concentrating most of the constraint capacity over the mutation
neighborhood.}
    \label{fig: iweights}
\end{figure}

\subsubsection{Information Weights}
\label{sec: information_weights}

Fig.~\ref{fig: iweights} shows the distribution of the information weights
$\alpha_i$ (see Eq.~\eqref{eq: information_weight}) for the \texttt{Astroalign} test suite.
The distribution is bimodal: a dominant peak around
$\alpha_i \approx 0.04$ groups the majority of tests, while a secondary
peak near $\alpha_i \approx 0.005$ corresponds to tests with reduced
marginal contribution. Two tests exhibit $\alpha_i = 0$.

A group-based analysis reveals an interpretable pattern.
Tests with $\alpha_i = 0$ cover degenerate cases of the algorithm in
which the core geometric matching procedure is not exercised.
Tests with $\alpha_i < 0.02$ verify interface contracts and type
conversions across different input formats. In both cases, their
mutation-killing capacity is largely absorbed by tests that exercise
the full processing pipeline.
The high-weight group concentrates the tests targeting the core
algorithm together with tests covering input formats not reached by
the standard cases.

These results reveal a fundamental limitation of code coverage as a
metric of test quality. Tests covering degenerate cases achieve
individual line coverage between 20\% and 27\%, comparable to the
26.5\% coverage obtained by interface verification tests.
Both groups contrast sharply with the tests of the core algorithm,
whose individual coverage exceeds 74\%.
Despite this similarity in coverage values, the former group exhibits
$\alpha_i = 0$ or very small values, while the latter concentrates the
largest weights in the distribution.
Code coverage is therefore unable to distinguish between a test whose
marginal contribution to entropy reduction is negligible and one that
imposes substantial constraints on the space of implementations;
$\alpha_i$ captures precisely this difference.

From a software engineering perspective, this distinction is critical.
Tests with large $\alpha_i$ are those that effectively sustain the
software specification: when executed in isolation, they eliminate a
significant fraction of mutants and reject most alternative
implementations.
If a modification alters the core logic, these are the tests that will
detect it.
By contrast, tests with small or null $\alpha_i$ impose weak
constraints: many program variants with different behavior can still
satisfy them.
A test suite composed predominantly of low-weight tests may achieve
high coverage while still providing a weak specification, allowing
incorrect implementations to pass undetected.


\section{Conclusions and Discussion}

In this work we introduced a formal definition of \textit{software entropy} grounded in the framework of statistical mechanics. The main contributions of this work can be summarized as follows:

\begin{itemize}
    \item We developed a rigorous theoretical framework for defining software entropy, consistent with the principles of statistical mechanics while remaining aligned with the intuitive notion of disorder frequently invoked by software developers.

    \item Within this framework, we identified concrete quantities that allow the restriction of program space induced by testing to be quantified. Based on these ideas, we proposed metrics for evaluating both code bases and test suites, including measures that capture the marginal contribution of individual tests to entropy reduction.

    \item We implemented a practical methodology, supported by the tool \texttt{Yagua}, that allows the local neighborhood of a program in the mutation graph to be explored and used to estimate entropy-related quantities.

    \item Finally, we demonstrated the application of these ideas in a real-world case study, showing how the proposed metrics behave in practice. In particular, we analyzed the entropy reduction trajectory of the test suite and the distribution of information weights across individual tests, revealing structural differences in their contribution to constraining program behavior.
\end{itemize}

\subsection{Conceptual Implications}

The formulation presented in this work provides a principled interpretation of software testing through the lens of statistical mechanics. In this framework, the set of tests plays the role of macroscopic constraints defining the observable behavior of a program, while concrete implementations correspond to microstates.

Software entropy therefore quantifies the uncertainty about the implementation given the behavioral constraints imposed by the tests.

Under this interpretation, the expansion of a test suite corresponds to a refinement of the macrostate: each additional test introduces a new constraint
that eliminates possible implementations compatible with the existing ones. As a consequence, the number of admissible microstates decreases and the entropy
is reduced. The empirical analysis presented in Section~\ref{sec: reducing_soft_entropy} illustrates this process explicitly, showing how successive additions to the test suite restrict the mutation neighborhood of the implemented program and lower the corresponding entropy upper bound.

This perspective provides a theoretical justification for a long-standing intuition among developers: improving the test suite reduces the uncertainty
about program behavior. In the language of statistical mechanics, testing can
be interpreted as a mechanism that progressively collapses the accessible region of program space.

The analysis of information weights presented in Section~\ref{sec: information_weights} further shows that not all tests contribute equally to this restriction process. While some tests impose strong semantic constraints on the implementation, others contribute only marginally because their mutation-killing capacity is largely redundant with that of other tests. The weights $\alpha_i$ provide a quantitative measure of this effect by identifying the mutants that are uniquely eliminated by each test. In this way, the proposed framework reveals an internal structure within the test suite that traditional metrics such as code coverage are unable to capture.
This distinction is particularly important from a software engineering perspective, since tests with large information weight effectively sustain the behavioral specification of the program, whereas suites dominated by low-weight tests may achieve high coverage while still leaving large regions of program space unconstrained.

\subsection{Methodological Considerations}

Several practical considerations arise when attempting to study software entropy empirically.

First, simulating the growth of a test suite for a fixed code base is relatively straightforward, as it only requires enabling additional tests. In contrast,
simulating the historical growth of a code base while preserving a consistent test suite is substantially more difficult. Modern software systems often exhibit
strong coupling between modules, which makes it challenging to remove parts of the code in order to reproduce earlier stages of development without breaking the system. This coupling problem limits the feasibility of reconstructing development trajectories directly from the present code base.

A natural alternative would be to analyze software repositories through version control systems. In principle, one could recover the entropy trajectory of a project by traversing its commit history and performing mutation analysis at each revision using the corresponding code and test suite. Such an approach would provide valuable insights into how software entropy evolves during real development processes.

However, implementing this strategy in a robust and automated manner presents significant engineering challenges. Mutation testing is computationally
expensive, and executing it across large commit histories would require careful management of execution environments, dependencies, and intermediate results. Moreover, many commits modify both the code and the tests simultaneously,
which complicates the interpretation of entropy changes. For these reasons, a systematic repository mining study was beyond the scope of the present work, whose primary goal is conceptual and theoretical.

\subsection{Future Directions}

The framework introduced here opens several directions for further research.

One promising avenue involves applying the methodology to other programming ecosystems. In particular, the Java ecosystem offers the state of the art mutation testing tools \citep{coles2016pit}, which may enable the analysis of larger and more diverse real-world projects. Porting the present methodology to such environments could provide a broader empirical basis for evaluating the proposed metrics.

Another important direction concerns the study of software evolution through repository mining. Although technically challenging, reconstructing entropy trajectories across project histories could provide valuable insights into how testing practices influence software quality during development \cite{hassan2008road}.

Finally, the exploration of program space need not be limited to traditional mutation operators. In the present work, mutation testing was used as a practical
mechanism for sampling nearby implementations in the mutation graph. However, recent advances in large language models suggest an alternative approach:
generating \emph{semantic mutants} that modify program behavior while still satisfying existing tests. Such mutants could allow the sampling of regions of program space that are far from the original implementation in the mutation graph but still compatible with the macrostate defined by the test suite. This strategy may provide a richer approximation of the accessible microstate space, leading to improved entropy estimates and a deeper analysis of the structural role played by individual tests in constraining program behavior \cite{wang2024comprehensive}.

Taken together, these directions suggest that the statistical-mechanical interpretation of software entropy introduced in this work may serve as a foundation for a broader quantitative theory of software testing and evolution.

\newpage

\backmatter



\bmhead{Acknowledgements}

This work was supported by CONICET and CONAE.
J.F. was supported by CONICET fellowships.
This work used computational resources from Instituto de Astronomía Teórica y Experimental (IATE) – Universidad Nacional de Córdoba (\url{https://iate.oac.uncor.edu/}).
This article has been revised using large language models (Claude,
ChatGPT, Gemini) in order to improve the clarity and correctness of the text.
All technical-scientific content remains property of the authors.

\section*{Declarations}

\subsection{Funding}
This work was supported by CONICET and CONAE.
J.F. was supported by a CONICET fellowship.

\subsection{Code and Data Availability}

The source code of \texttt{Yagua} used in this work is available at
\url{https://github.com/leliel12/yagua/tree/paper} (\texttt{paper} branch)
and can be installed directly via \texttt{pip}:

\begin{verbatim}
$ pip install
    https://github.com/leliel12/yagua/archive/refs/heads/paper.zip
\end{verbatim}

The software is released under the BSD 3-Clause License \citep{bsd3}.
The repository also contains the results of running \texttt{Yagua} against
\texttt{Astroalign}, together with a Jupyter notebook implementing the
algorithms used to generate the figures presented in this work.

\subsection{Conflict of interest/Competing interests}
The authors declare that they have no competing interests.

\subsection{Ethics approval and consent to participate}
This study does not involve human participants, human data, or animals. No ethical approval was required.

\subsection{Consent for publication}
Not applicable.

\subsection{Author contributions}
J.F. developed the theoretical framework and formalization of software entropy.
J.B.C. implemented the computational tools and performed the mutation analysis
and simulations. Both authors contributed to the design of the study, the
interpretation of the results, and the writing of the manuscript.

\begin{appendices}

\section{On Other Types of Testing}\label{appendix: types_of_testing}


\subsection{Genetic Testing as Semantic Constraint Discovery}

Search-based (genetic) testing explores input space and often discovers behavioral
properties that the implemented program appears to satisfy consistently, even when these
were never specified. Each discovered invariant may be viewed as a new semantic restriction
$t_{m+1}$, yielding a refined macrostate
\[
\mathbb{P}'=\{\,p\in\mathbb{P} : t_i(p)=1,\ i=1,\dots,m+1\,\}.
\]
In this sense, genetic testing complements mutation testing: while mutation testing exposes alternate microstates that satisfy the current macrostate (that we might want to exclude from $\mathbb{P}$ through additional tests), genetic testing uncovers latent constraints that characterize the actual microstate $p_{\mathrm{impl}}$. Both mechanisms suggest directions in which the macrostate may be tightened and both contribute to entropy reduction, albeit from opposite viewpoints.

\subsection{Fuzzy and Property-Based Testing}

Property-based tests need not impose crisp boolean restrictions. Let
$\tilde t_i\colon\mathbb{V}\to[0,1]$ denote the empirical probability that program $p$
satisfies property $i$ under randomized inputs. These values define a weighting
\[
w(p) := \prod_{i=1}^m \tilde t_i(p),
\qquad
\nu(p) := \frac{w(p)}{\sum_{q\in\mathbb{V}} w(q)},
\]
and the entropy of this fuzzy macrostate is the Gibbs--Shannon functional
\[
S[\nu] = -\sum_{p\in\mathbb{V}} \nu(p)\,\log \nu(p).
\]
The crisp case corresponds to $\tilde t_i\in\{0,1\}$. This formulation accommodates
testing regimes where properties are evaluated stochastically or only approximately, and
connects naturally with non-uniform ensembles in non-equilibrium statistical mechanics.

\subsection{Additional Testing Paradigms}

Other testing approaches fit naturally into this formalism. Differential testing compares
independently developed implementations and restricts $\mathbb{P}$ to their behavioural
intersection. Static analysis tools, although mostly syntactic, may also be interpreted as
imposing admissibility constraints on $\mathbb{V}$ (analogous to bounding $L_{code}$),
thereby reducing the configuration space prior to semantic restriction. Specification
mining or property-driven search impose further semantic constraints without explicitly
generating alternate microstates.

\section{On Local Test-Suite Metrics} \label{appendix: local_metrics}
\subsection{Software Entropy Density and its Local Proxy}

For a fixed code length $L_{\text{code}}$, the theoretical Software Entropy Density (SED)
associated with a macrostate is
\[
    \text{SED} := \frac{S}{L_{\text{code}}}
    = \frac{-\log W}{L_{\text{code}}},
\]
where $W = |\mathbb{P}|$ is the number of admissible microstates. This quantity measures the
average logarithmic ``freedom'' per unit of code length. In principle, a smaller SED
indicates that the behavioral constraints encoded in the tests describe the program more
precisely, leaving fewer admissible implementations.

In practice, however, the global cardinality $|\mathbb{P}|$ is intractable: even ignoring
semantic undecidability issues, its computation would require enumerating all syntactically
valid programs up to length $L_{\text{code}}$. What is accessible experimentally is a
local approximation obtained from mutation analysis. Given a family
$\mathcal{F} = \{T_0,\dots,T_m\}$ of increasingly restrictive test sets and their associated
local feasible regions $M_0(p_{\mathrm{impl}}),\dots,M_m(p_{\mathrm{impl}})$, we define the
\emph{local Software Entropy Density} as
\[
  \text{SED}_{\mathrm{loc}}(p_{\mathrm{impl}})
  := \frac{\log|M_0(p_{\mathrm{impl}})| - \log|M_m(p_{\mathrm{impl}})|}
           {L_{\text{code}}}.
\]
This quantity measures the reduction in the local entropy upper-bound induced by the full
test suite, normalized by code length. It probes only a neighborhood of
$p_{\mathrm{impl}}$, yet provides a meaningful reference scale when comparing different test
suites, different versions of the same project, or different testing strategies. It should
be interpreted explicitly as a \emph{local} upper-bound contribution, not as an estimator of
the global SED, but it remains a useful practical proxy for the degree to which tests
constrain nearby microstates.

\subsection{Macrostate Tightness Indexes}
In addition to mutation score, two complementary metrics quantify how uniformly a test
suite restricts program space. For each test $t_i$, let $K_i$ be the set of mutants only killed by $t_i$. The \emph{tightness ratio}
\[
MTI_1 := \frac{|\{\,i:\,K_i\neq\varnothing\,\}|}{m}
\]
measures the proportion of non-redundant (or more precisely, non-overlapping) constraints. The \emph{information-weighted
tightness}

\begin{equation*}
MTI_2 := \frac{-\sum_{i=1}^m \alpha_i \log \alpha_i}{\log m},
\qquad
\alpha_i := \frac{|K_i|}{\sum_{j=1}^m |K_j|},
\end{equation*}
quantifies how evenly constraint power is distributed across the test suite. Moreover, high $MTI_2$ might be linked to high test modularity. Both metrics capture aspects of macrostate refinement that mutation score alone does not reflect.

If tests were highly overlapping, that is, almost all mutants are eliminated by more than one test, alternative metrics could be calculated by taking $K_i$ to be the set of mutants killed by $t_i$ when run alone.

\end{appendices}


\bibliography{sn-bibliography}

\end{document}